# Experimental Study of the Acoustic Field Generated by a 50 MeV Electron Beam in Water

V. B. Bychkov[1], V. S. Demidov[2], E. V. Demidova[2], A. N. Ermakov[3], O. D. Ershova[3], B. S. Ishkhanov[3], V. P. Maslyany[1], A. Yu. Sokolov[2], N. A. Khaldeeva[2]

1. All-Russian Scientific Research Institute of Physical-Technical and Radiotechnical Measurements, Mendeleevo, Moskovskaya obl., Russian Federation
2. Alikhanov Institute for Theoretical and Experimental Physics, Moscow, Russian Federation
3. Skobeltsyn Institute of Nuclear Physics, Lomonosov Moscow State University, Moscow, Russian Federation

At the MSU SINP electron accelerator, a space-time dependence of the acoustic pressure generated in water by an electron beam of 50 MeV energy was obtained. Measurements were carried out in 100 points located along the line parallel to the beam axis at the distance of 6.5 cm from the axis. At a two-dimensional diagram (distance-time) two signal tracks were observed from two sound sources: a cylindrical acoustic antenna generated by the electron beam, and an area of the beam entrance cap which divides the water medium from the air.

## Introduction

The effect of acoustic field origination due to the passage of ionizing particles through matter is interesting in connection with the idea to use this effect for measuring the spectra of ultra-high energy neutrinos in natural water basins. Obtaining precise experimental data in this field will help to answer the questions such as what are the mechanisms of ultra-high energy cosmic rays generation and what astrophysical objects could be their sources, and will also enable to check the validity of different cosmological models of the origin of the Universe. At the moment a number of large underwater neutrino detectors are developed within international collaborations. In most of them the detection of Cherenkov light from charged particles produced by neutrino interactions in water is used as a primary method. In some experiments (ANTARES, NEMO, NT-200, etc.) a complementary method based on the registration of the acoustic radiation generated due to the passage of charged particles through matter is also developed. It is well-known that neutrino interaction with matter results in the production of charged leptons (electrons and muons), which then initiate the development of hadron-electromagnetic cascades. Dimensions of these cascades reach several meters in length and tens of meters across. According to Askarian [1] hypothesis, confirmed by experimental data systematized in the monography [2], the passage of charged particles through matter is accompanied by mechanical oscillations of the medium, a spectrum of which in prospective experimental conditions lies within 1-200 kHz frequency band. It is assumed that by recording the acoustic pulse from the hadron-electromagnetic shower and measuring its acoustic pressure it is possible to estimate the energy of the neutrino as a shower source. Moreover, the determination of the shower development direction and its spatial position yields the direction of the primary neutrino.

At the stage of planning the experiments it is significant to obtain expected response values of the detection equipment. In the experimental investigation of the phenomena acoustic signals from cascade showers are simulated by signals generated in small-sized tanks due to absorption of low-energy particle beams in water. The experiments are carried out at the accelerators with available intensities high enough to approach the energy deposition values expected in neutrino experiments.

Properties of the acoustic signals have been mainly studied with proton beams [3 - 6]. Few experiments were conducted with electrons more than 20 years ago [2, 7]. In these experiments five signals were registered at the distances from 3 to 18 cm from the beam axis.

In the present work a space-time dependence of the acoustic field, the source of which is a 50 MeV electron beam, is measured and its structure is analyzed. According to the common notions of acoustic signal generation mechanisms, electron-photon cascades produced by electrons traversing water medium form an acoustic antenna of quasi-cylindrical shape with alternating diameter. The aim of the experiment was the registration of acoustic signals radiated by the antenna in a number of points at the distances comparable to the length of the antenna and significantly exceeding its transverse dimensions.

**Experimental equipment**

The experiment was conducted at the MSU SINP pulse race-track microtron RTM70. The energy of electrons in the experiment was 50 MeV, beam spill time – 8 μs, pulse frequency – 10 Hz. The beam cross-section shape was quasi-elliptical (with vertical dimensions of about 5 mm and horizontal dimensions of about 2.5 mm). The beam current was controlled via a beam current transformer sensor and had an average value of 2 mA, which corresponds to an average intensity of about $10^{11}$ particles per beam spill. A total energy deposition per spill was of the order of $5·10^{18}$ eV.

As a target, in which the investigated ultrasonic field was generated, a tank filled with distilled water was used. The tank was made of acrylic plastic and was reinforced and sealed at side joints. It had a parallelepiped shape with the dimensions of 50.8 x 52.3 x 94.5 cm. The experimental set up with all dimensions needed for data analysis is shown in fig. 1. The level of water in the tank was 390 mm, its average temperature was equal to 20.5 °C. A possible local temperature increase due to the beam passage was not controlled. The tank dimensions were large enough to distinguish in process of the analysis between a direct signal from the beam and signals reflected from the walls. The electron beam was injected into the center of the measurement volume through a duralumin tube of 59 mm diameter, 460 mm length and 1.5 mm wall thickness, which was inserted in a lateral face of the tank and was closed with a teflon cap of 2 mm thickness.

The measurements were performed using a broadband (up to 160 kHz) highly sensitive (> 1 mV/Pa) measuring hydrophone, which was designed by VNIIFTRI specialists for radiation acoustic measurements. A piezoelectric ceramic with tangential polarization was utilized as a sensor.

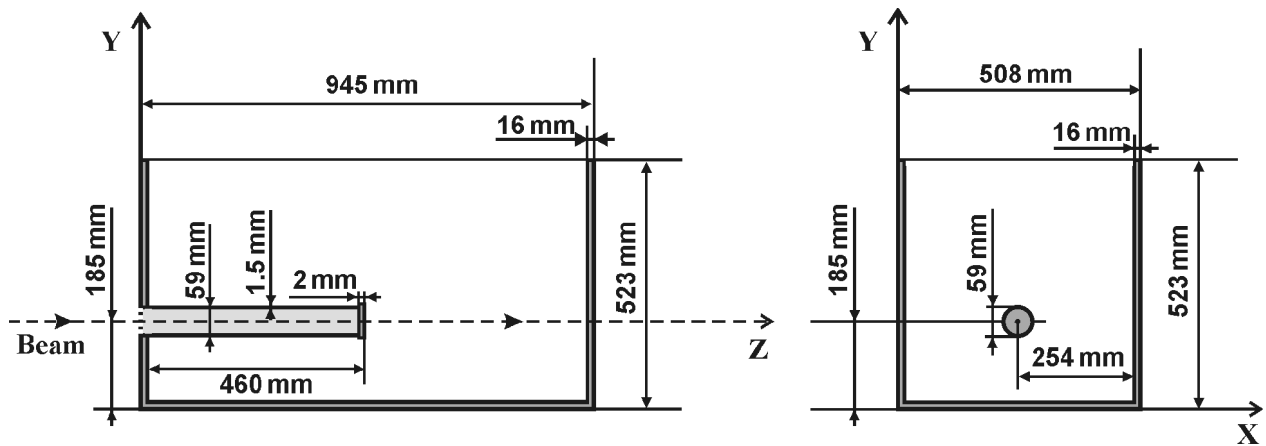

*Fig. 1.* Experimental set up.

A schematic diagram of the experimental set up is presented in fig. 2. The hydrophone was connected through its own preamplifier and also two amplifiers giving the amplification of 50 and 10 dB in the band of 20 Hz – 200 kHz and 10 –100 kHz respectively.

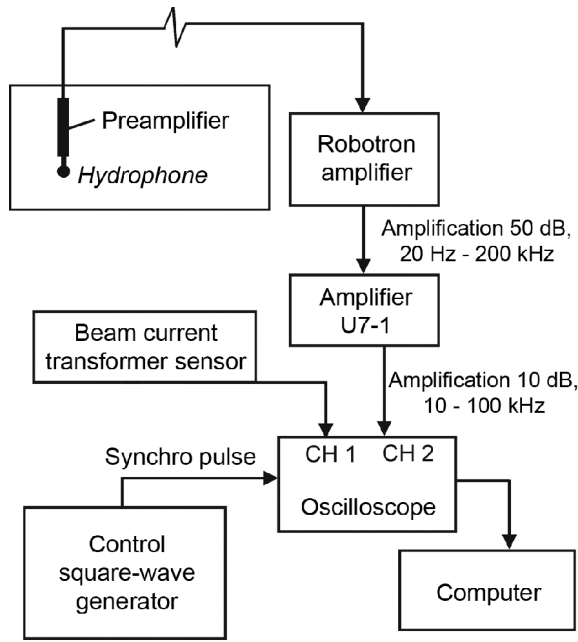

*Fig 2.* Schematic diagram of the experiment.

The data was read out using the TEKTRONIX TDS 3032 2-channel oscilloscope operated by a personal computer. The signal observation time was 1 ms, each measurement contained 10000 readings by 0.1 μs each. The oscilloscope was triggered by a synchronizing pulse from a generator controlling the accelerator performance and anticipated the time of the beam arrival to the target by approximately 4 μs. A signal from the beam current transformer was fed into one oscilloscope channel, the signal from the hydrophone – to another. Oscillograms were recorded to the computer disk in a native oscilloscope file format.

The oscillograms of the acoustic signals were recorded in 100 points along the straight linear track, parallel to the electron beam axis and located in the horizontal plane with the axis. The distance between the beam axis and the track was $X_0 = 6.5$ cm.

The hydrophone was moved along the track using a special electromechanical remote-control scanner designed by V.I. Albul [5]. The scanner step was equal to 4.45 mm. A starting point $Z$ coordinate coincided with the origin of the acoustic antenna.

**Results and discussion**

In fig. 3 a fragment of a signal from the hydrophone recorded at the distance of 8 cm from the scanning start point is shown with a dash line. The time sweep from the moment of the sync pulse arrival is represented by the abscissa axis, the voltage value in Volts – by the ordinate axis. The beginning and the end time of the beam spill (4–12 μs) is marked by arrows. The acoustic signal generated by the beam is located within a time interval from 34 to 85 μs, and it has a bipolar shape. At the time $t = 18$ μs an electromagnetic noise pulse related with the operation of the accelerator is observed. Its amplitude exceeds the amplitude of the studied signal by several times and impedes the analysis possibility at short times ($< 36$ μs) and, consequently, at short distances between the sound source and the detector.

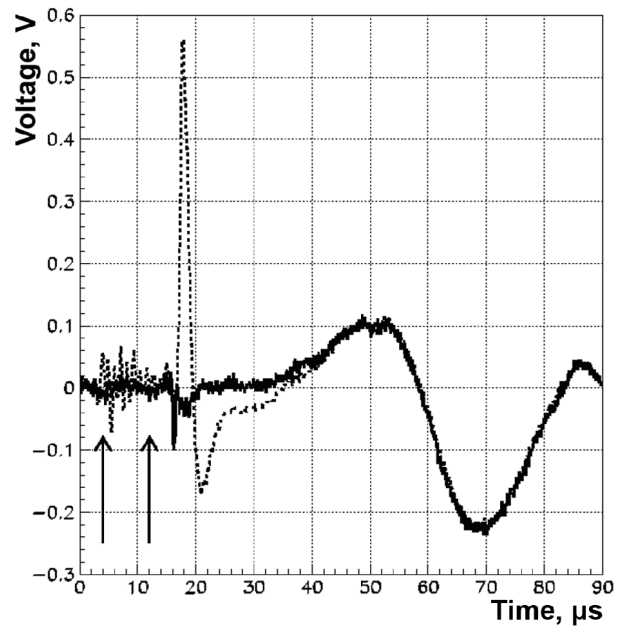

*Fig 3.* Signal from the hydrophone at $Z = 8$ cm from the origin of the acoustic antenna and $X_0 = 6,5$ cm from its axis. A dotted line shows the initial signal, a solid line – the signal after subtracting the noise pulse. Arrows mark the beam spill time.

It is possible to significantly reduce the noise pulse influence by subtracting it from the recorded signal. The noise pulse was measured while the accelerator was operating in the absence of the electron beam. It determines the signal shape in a time interval 18–36 μs. The subtraction result is presented in fig. 3 with a solid line. The figure implies that the noise pulse subtraction procedure allows to suppress the spurious signals influence by at least 7 times at the times less than 25 μs and to decrease the lower boundary of the studied interval until 2-5 μs from the signal beginning, which is important for a more detailed study of radiation acoustic waves generation mechanisms. Results presented below were obtained using the above-mentioned procedure.

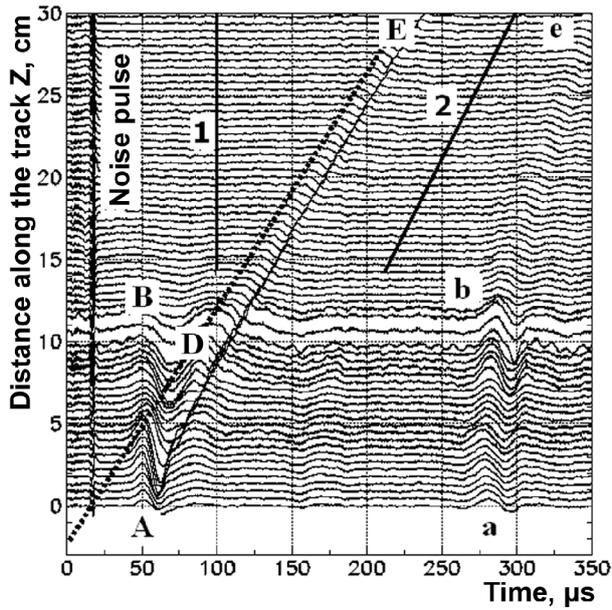

*Fig 4.* Space-time structure of the acoustic field generated by an electron beam in water.

Time dependencies of ultrasonic pulses recorded in 66 points along the measurement track are presented in fig. 4. The signals are normalized to 1 mA beam current. The ordinate axis represents distances along the track in cm, which enable to determine the coordinates of the measurement points inside the tank.

Acoustic signals form three paths in the (*t-Z*) plane, two of which (AB and DE) have the beam propagation area as a source, and the third (abe) represents the reflections of the signals produced by the beam from the bottom of the tank. The AB path formed by ridges of alternating amplitude corresponds to a first half-wave of the acoustic signal (compression half-wave) from the nearest point of the radiating acoustic antenna. The path is almost parallel to the distance axis, because the propagation time of this signal from the source to the detector is the same, accurate within variation of electron-photon cascade transverse dimensions.

Signals which form the DE path begin with a rarefaction half-wave. Considering the arrival time of these signals, the shape (straight line) and the direction of the path, it can be concluded that the source of these signals is located in the area of the cap which divides the water medium from the air. Similar signals were earlier detected in experiments with electron [7] and proton [6, 8] beams. A dotted line in the figure 4 is a result of approximation of the points corresponding to the beginning of acoustic signals by a linear dependence

$$R_i = V t_i + r, \qquad (1)$$

where $i$ is a number of the track point with coordinate $Z_i$, in which the signal was recorded (the Z coordinate zero is at the beginning of the scanner track), $t_i$ – time of the signal arrival to the hydrophone, $R_i = \sqrt{X_0^2 + Z_i^2}$ – distance from the center of the cap to the hydrophone. The approximation was carried out for the track segment from 13 to 37.5 cm. Numerical values of the speed of sound $V$ and the distance $r$ from the sound source at the starting point of the track were determined as a result of approximation.

To determine the signal arrival time $t_i$ the following algorithm was used. In each point $i$ the minimum of the acoustic signal was determined. The interval for the minimum search was in the zero-order approximation limited by a time interval in which the analyzed signal was the only. Lower and upper limits of this interval are presented in fig. 4 by lines 1 and 2, respectively. In fig. 5 a fragment of the signal recorded at $Z = 15.6$ cm is presented; the found minimum is marked by an arrow denoted by $t_{min}$. Then, using the least square technique in an interval of $(t_{min} - 120\ \mu s < t < t_{min} + 5\ \mu s)$ parameters of an approximating Gaussian were determined. Its average $t_G$ and standard deviation $\sigma$ were used to calculate the signal arrival time $t_0 = t_G - 2\sigma$. It is shown in fig. 5 by an upward arrow. Thus, for each of 57 points the times $t_{0i}$ and distances $R_i$ were known. These pairs of values are plotted in fig. 6. The straight line is a result of approximation of points by the linear dependence (1). The determined speed of sound value at this track appeared to be equal to $V = (1435 \pm 12)$ m/sec. This value can be compared to the speed of sound value which was measured during a calibration experiment carried out in the same tank with a piezoelectric radiator as a sound source and was determined to be $(1461 \pm 15)$ m/sec. The values agree within two standard deviations, and small discrepancies might be taking place because of finite dimensions of the radiator and the sound detector.

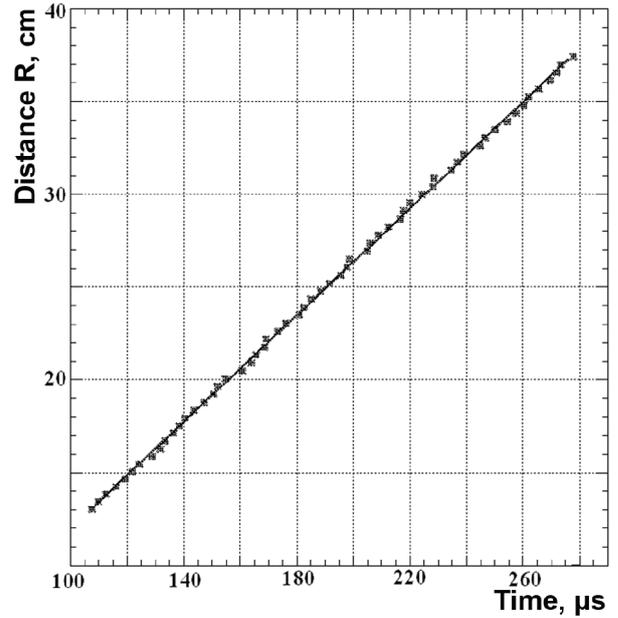

Fig 6. Time dependence of the distance between the sound detector and the center of the cap which divides the water from the air. A straight line is a result of the approximation of experimental points.

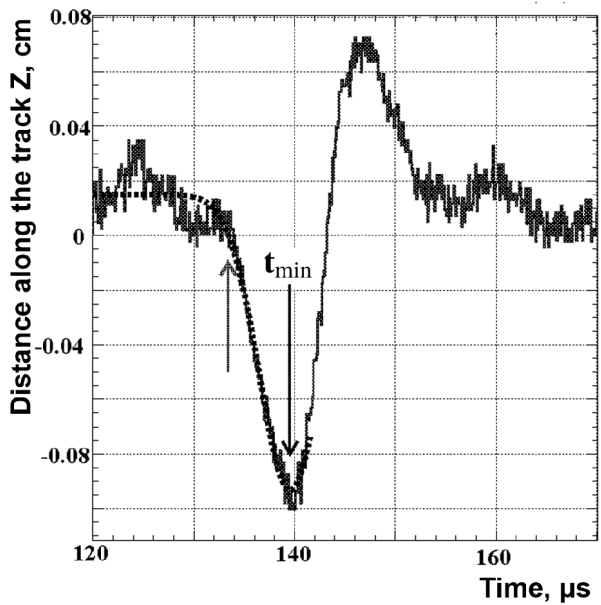

Fig 5. An illustration to the signal time arrival determination algorithm. A dotted line shows an approximation of a signal recorded at $Z = 15.6$ cm from the origin of the electron-photon cascade by a Gaussian.

As for the calculated value of coefficient $r = (-2.50 \pm 0.05)$ cm, its significant deviation from $X_0$ at $t = 0$ probably is the evidence of not identical speed of sound in undisturbed medium (far from the source) and in the medium where signals from two the sound sources overlap. One of the reasons of the sound speed increase near the electron beam could be a local temperature increase due to the beam influence. The change of the speed of sound at small distances from the beam can be observed qualitatively in fig. 4, where at times $t > 90\ \mu s$ the signals follow the approximation line, but at smaller times deviate from it.

### Conclusions

As a result of the experiment, for the first time a detailed space-time dependence of the acoustic field, generated in water by an intensive electron beam, was obtained at the track parallel to the beam axis. The experimental method prevented the influence of reflected signals in the investigated area, which was about 30 cm and exceeded longitudinal dimensions of the acoustic antenna, generated due to the beam propagation in water, by approximately 3 times. The utilized algorithm of data processing and the way of presenting the results enabled to distinguish confidently signals from the cylindrical acoustic antenna itself and the signals from the area of the cap through which the beam enters the tank. The analysis of the latter signals propagation speed implies that there is a possible local increase of the speed of sound in the area of interference of the two discussed signals. The reason for that is supposed to be studied in further experiments.